\documentclass[nofootinbib,prl,floatfix,preprintnumbers,
amsmath,amssymb,
twocolumn]{revtex4}
\usepackage{epsfig}

\def\be{\begin{equation}}
\def\ee{\end{equation}}
\def\bea{\begin{eqnarray}}
\def\eea{\end{eqnarray}}
\begin{document}
\title{Quantum Phases of a Vortex String}
\author{Roberto Auzzi and S. Prem Kumar}
\email[]{r.auzzi, s.p.kumar@swansea.ac.uk}
\affiliation{
Department of Physics, Swansea University, Singleton Park, Swansea, 
SA2 8PP, U.K.}
\begin{abstract}
We argue that the world-sheet dynamics of magnetic $k$-strings in the
Higgs phase of the mass-deformed ${\cal N}=4$ theory, is controlled by
a bosonic $O(3)$ sigma model with anisotropy and a topological $\theta$
term. The theory interpolates 
between a massless $O(2)$ symmetric regime, a massive $O(3)$ symmetric
phase and another massive phase with a spontaneously broken ${\mathbb Z}_2$
symmetry. The first two phases are separated by a Kosterlitz-Thouless
transition. When $\theta=\pi$, the $O(3)$ symmetric phase flows to an
interacting fixed point; sigma model kinks and their
dyonic partners become degenerate, mirroring the behaviour
of monopoles in the parent gauge theory. 
This leads to the identification of the kinks with monopoles confined
on the string. 
\end{abstract}
\maketitle

{\it Introduction.--} The formation and 
dynamics of colour flux tubes is of fundamental importance to the physics of
gauge theories that exhibit confinement. The flux tubes or ``QCD
strings'' have their own intrinsic dynamics and degrees of
freedom. This raises an intriguing question, namely, what is the
relation between the  world-sheet degrees of freedom of the confining
strings and the
underlying gauge theory physics. This is a
difficult problem in general, but can become tractable 
if the gauge theory has global
symmetries that yield light, internal modes on the world-sheet with
non-trivial dynamics.
In certain examples with such internal symmetries and 
adequate amounts of supersymmetry, the
connection between the supersymmetric 
 dynamics of magnetic flux tubes or vortices,  
and their parent four dimensional field theories can
be demonstrated beautifully \cite{Shifman:2004dr}. 
In this letter we reveal a similar connection for 
a purely bosonic sigma model on a flux tube that resides 
in an ${\cal N}=1$
supersymmetric gauge theory. The novel feature of our example is, in
the absence of supersymmetry, the appearance of 
a rich quantum phase structure as a function
of its parameters, some features of which reflect the 4D physics.

The gauge theory in question is a mass deformation of
${\cal N}=4$ SUSY Yang-Mills, the so-called ${\cal N}=1^*$ theory
realized in its Higgs phase, and the vortex strings in this vacuum are
the magnetic versions of confining $k$-strings. 
The ${\cal N}=4$ SUSY gauge theory can be viewed
as the theory of an ${\cal N}=1$ SUSY
vector multiplet ${\cal W}^\alpha$, and three adjoint chiral
multiplets $\Phi_a$, $(a=1,2,3)$. Suitable deformations of this
theory can lead to rich infrared physics. One such 
deformation is the ${\cal N}=1^*$ theory, corresponding to 
non-zero masses for the three chiral multiplets. 
The resulting ${\cal N}=1$ theory, with gauge group
$SU(N)$, has a large number of vacuum states, and classically these
are in one-to-one correspondence with the partitions of $N$ into
integers \cite{dw}. Of particular interest are vacua with a mass gap.
Interestingly, every
possible massive phase, including Higgs and confining, of an
$SU(N)$ gauge theory with adjoint matter, 
is realized by one of the vacua. 
The ${\cal N}=1^*$ theory has a classical superpotential,
\be
W= 
{\rm Tr}
\left(
\Phi_1[\Phi_2,\Phi_3]+ \tfrac{1}{2}\sum_{a=1}^3
  m_a \,\Phi_a^2
\right),
\ee
where the cubic term is the superpotential of the conformal ${\cal
  N}=4$ theory. Classical ground states are determined by the
F-term equations ${\partial W}/{\partial \Phi_a}=0$,
which are solved by $N$-dimensional representations of the $SU(2)$
algebra. The irreducible representation yields the ``Higgs vacuum''
\be
\Phi_{1,2}= i \sqrt{m_{1,2} m_3} \;\left[J_{1,2}\right]_{N\times N},\,\,\,\label{vevs}
\Phi_3 =i \sqrt{m_1 m_2}\;\left[J_3\right]_{N\times N},\nonumber
\ee
where $J_a$ are the $N$-dimensional generators of $SU(2)$. The
gauge symmetry is completely broken in this vacuum and magnetic
degrees of freedom are confined. 
Amongst the many other massive vacua, there is also one with $\Phi_a=0$, where 
the gauge group is completely unbroken and the quantum dynamics
confines electric degrees of freedom.
The different massive phases of the theory are exchanged and permuted by the
action of the $SL(2,{\mathbb Z})$, Montonen-Olive duality group of the
parent ${\cal N}=4$ theory. In particular, the electric-magnetic
duality or S-duality: $g_{YM}\rightarrow 1/g_{YM}$, 
swaps the Higgs and confining phases. Consequently, for a
fixed $N$, the physics of the confining vacuum at strong coupling
$g_{YM}\gg 1$ is well described by the Higgs vacuum at weak coupling, 
$g_{YM}\ll 1$.
 
{\it Vortices in ${\cal N}=1^*$ theory.--}
The Higgs vacuum at weak coupling admits classical vortex
solutions \cite{mamayung, us}. These solutions carry a discrete
magnetic flux, taking values in $\pi_1[SU(N)/{\mathbb Z}_N]\simeq{\mathbb Z}_N$. Solutions 
carrying $k$ units of ${\mathbb Z}_N$ flux, 
the so-called magnetic $k$-strings (S-dual to confining $k$-strings)
were studied extensively in \cite{us}, specifically when the mass
deformation parameters were equal: $m_a=m$. In this situation,
the gauge theory has a global $O(3)$ symmetry, under which $\Phi_a$
transform as a triplet, which is broken by
the VEVs \eqref{vevs} in the Higgs phase. However, a combination of 
global colour rotations and the broken $O(3)$ symmetry, leaves the Higgs
VEVs invariant and a ``colour-flavour locked'' symmetry, $O(3)_{C+F}$ remains.

The configurations of interest 
have the adjoint scalars $\Phi_a$ approaching their VEVs in the Higgs
vacuum asymptotically, along with a phase winding around the vortex. 
This phase rotation corresponds to a gauge transformation (at infinity) 
which is single-valued in $SU(N)/{\mathbb Z}_{N}$ and is generated by
\be 
Y_k= \tfrac{1}{N} \;{\rm Diag} 
(\underbrace{k,\cdots,k}_{N-k\,{\rm times}} , - (N-k),\cdots, -(N-k) ), 
\ee
so that the resulting chromomagnetic flux is proportional to $Y_k$ and 
$\exp(i\oint A\big|_{r\to\infty})=\exp({2\pi i k/N})$. The flux
picks out a specific direction in the colour-flavour space and the associated
string is truly non-Abelian, as we illustrate with the example for
$SU(2)$ gauge group. For the $SU(2)$ gauge theory, the ${\mathbb Z}_2$
vortex ansatz \cite{mamayung} in singular gauge reads,
\bea
&&\Phi_1= i m\, \psi(r) \, \tau_1\,;\,\, \Phi_2= i m \,\psi(r)
\,\tau_2\,;\,\,\Phi_3 = i m \kappa(r)\,\tau_3,\nonumber\\
&&\vec A= \tfrac{f(r)}{r}\,Y_1\,\hat\phi\,,\;\; Y_1=\tau_3.
\label{z2}
\eea
Here $\tau_a$ are generators of $SU(2)$, and $\psi,\kappa, f$ are the
vortex profile functions which can be solved for
numerically. Crucially, the solution preserves a $U(1)$ subgroup
of the 
$O(3)_{C+F}$, corresponding to rotations in  the
$1$-$2$ plane. A generic colour-flavour rotation will change the
internal orientation of the non-Abelian flux and generate a family of
such solutions. There is therefore an $S^2 \simeq SO(3)/U(1)$ moduli
space of solutions. This picture and the solutions can be generalized
to the case of a general $k$-string for $SU(N)$ gauge group \cite{us}
and in all cases there is an $S^2$ moduli space of solutions.

{\it Vortex Sigma model.--}
By making the internal zero modes depend slowly on the vortex world-sheet
coordinates, and plugging in the associated ansatz into the gauge
theory action, it is possible to systematically derive the effective 
$1+1$-dimensional sigma model on the world-sheet \cite{mamayung,us}. The vortex
solutions break all four supercharges of the theory, and therefore we
do not expect fermionic internal zero modes (apart from the four
generated by the broken supercharges); an explicit search 
supports this. We thus obtain a bosonic sigma model with an
$S^2\simeq {\bf CP}^1$ target space, which we may conveniently view as
an $O(3)$ nonlinear sigma model:
\bea
&{\cal L}_\sigma= \tfrac{1}{2g_\sigma^2}
(\partial_\alpha \vec n)^2 +
\tfrac{\theta}{4\pi}\,\epsilon^{\alpha\beta}\vec n\cdot
(\partial_\alpha\vec n\times \partial_\beta \vec n),\\\nonumber
& \vec n\cdot \vec n=1\,;\qquad \theta= k(N-k)\theta_{YM}.
\label{sigmamodel}
\eea
Here $\vec n \equiv (n_1,n_2,n_3)$ is an $O(3)$ unit vector
parametrizing the internal orientation of the non-Abelian flux. $\theta
_{YM}$ is the vacuum angle of the 4D gauge theory, and  the bare coupling 
$g_\sigma=g_{YM}/C_{k,N} \ll 1$ with $C_{k,N}$ a numerically determined
function of $k$ and $N$. The $O(3)$ sigma model has instantons and the
associated topological $\theta$-term is fed in from the the gauge theory
vacuum angle through the simple but non-trivial relation above. This
relation has been obtained explicitly by using both semiclassical
methods and the large-$N$ gravity dual picture of ${\cal N}=1^*$ theory
\cite{us}. The sigma model description of the dynamics is
valid on length scales larger than the vortex thickness which can be
estimated  to be $\Lambda^{-1}\sim (m \sqrt N)^{-1}$. (This follows
from the connection between ${\cal N}=1^*$ vortices and noncommutative
instantons on $S^2 \times {\mathbb R}^2$.) 

It is well known that the $O(3)$ model is asymptotically free, and for
generic $\theta$, it has a mass gap. The dynamical scale of the theory
is $\Lambda_\sigma\sim  \Lambda \,\exp({-2 \pi/g_\sigma^2})$, and its
spectrum consists  of a single massive triplet. When $\theta=\pi$
however, the theory is known to flow to a $c=1$ conformal fixed point
described by the $SU(2)$ Wess-Zumino-Witten model at level one
\cite{haldane}. 

We now wish to consider what happens to the sigma model when we move away
from the $O(3)$-symmetric limit. In the four dimensional gauge theory
it is natural to consider the case where $m_1=m_2=m$ whilst $m_3 \neq
m$. Then the $O(3)$ global symmetry is explicitly broken to $O(2)$,
corresponding to rotations in the $\Phi_1$-$\Phi_2$ plane.
When $m_3 \ll m$, the theory can be viewed as softly broken ${\cal
N} =2^*$ gauge theory and in the opposite regime $m_3 \gg m$ it flows
toward an ${\cal N}=1$ superconformal field theory with two adjoint chiral
multiplets and a quartic superpotential. 

From the point of view of the $k$-string sigma model, it makes sense
to consider only small deviations from the $O(3)$ symmetric
situation, so that it is still meaningful to regard the internal
orientation as a quasi-modulus. To this end we introduce the deviation 
$\Delta\equiv m_3^2-m^2$. As long as $|\Delta/m^2| \ll 1$, the $O(3)$
breaking will manifest itself as a deformation of the sigma model
above.  We can then explicitly
compute this deformation potential at linear order in $\Delta$. At
this lowest order,   
it is consistent to take the unmodified vortex profiles
(e.g. \eqref{z2} for $SU(2)$), perform a generic colour-flavor
rotation and substitute into the 4D action with $m_3\neq m$, to
obtain the effective deformed sigma model in 2D, and we find
\be
{\cal L}_\sigma \to {\cal L}_\sigma - A_{k,N}\,\Delta\,(n_3)^2,
\ee
where $A_{k,N}>0$, is a constant that can only be determined
numerically for each $k$ and $N$. At higher order in $\Delta$, the
vortex solution itself will be modified and the potential will be 
complicated. However, the lowest order contribution is already
interesting. The key point here is that, depending
on the sign (and magnitude) of $\Delta$, the sigma model is in one of three
possible phases. Let us discuss these in succession. 
\begin{figure}[h]
\begin{center}
\epsfig{file=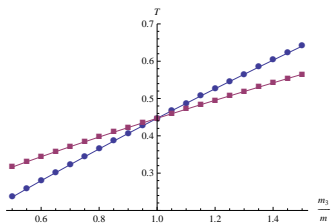, width =1.7in}
\end{center}
\caption{ Tensions of the $n_3=\pm 1$ vortex (round markers) and of
  the "equatorial" vortex (square markers) for different values of
  $m_3/m$ and $N_c=2$.} 
\end{figure}
Classically, when $\Delta <0$, we expect that $n_3=\pm 1$ are the
vacua, while for $\Delta > 0$, the equator of the target sphere
becomes the vacuum manifold. This picture is confirmed in Fig.1, by
computing numerically, the tensions of the {\em exact} vortex solutions oriented
along two different directions.

{\bf {\underline{$\Delta > \Lambda_\sigma^2$}}}: When $\Delta$ is much
larger than the dynamical scale $\Lambda_\sigma^2$ of the undeformed
$O(3)$ theory (still ensuring $\Delta/m^2\ll 1$ or $m_3\sim m$), 
the effective coupling $g_\sigma(|\Delta|)$ is weak
and the classical potential forces $n_3$ to vanish, keeping the
orientation in the 1-2 plane. The resulting $O(2)$ symmetry is
not broken due to Coleman's theorem, but there is a massless free
boson which is the angular degree of freedom. This model 
contains vortex-instantons which are suppressed and dilute for large $\Delta$,
but as $\Delta$ is decreased, and the effective coupling $g_\sigma$
increases, they become important. The  ``vortices inside the vortex''
come in two varieties. One that circulates around the equator at
infinity and moves off at its core to the north pole, while the second
kind moves to the south pole at the core. The topological  
charges  
of these two types of vortex instantons are $\pm 1/2$ and so they are
merons of the $O(3)$ model \cite{affleck}. As $\Delta$ is decreased so
that $\Delta\sim \Lambda_\sigma^2$,
there will be a critical value of the effective coupling
$g_\sigma(|\Delta|)$, at which the vortices will condense, following
a Kosterlitz-Thouless transition. Thus, a mass gap is generated by
meron condensation and the theory enters the massive $O(3)$ symmetric
regime. At $\theta=\pi$, this mechanism fails and the model remains
massless due to a cancellation between merons of 
positive and negative topological, meron charge 
\cite{affleck}. The Coulomb
gas of the two kinds of vortices can be mapped to a sine-Gordon model
for general $\theta$, with the action ${\cal L}\sim 
{g_\sigma^2}(\partial \varphi)^2/2-2 \zeta 
\cos\tfrac{\theta}{2}\cos(\varphi)$ ($\zeta$ is the vortex
fugacity). At $\theta =\pi$ this theory is massless.   

{\bf{\underline{$\Delta <0$ and $|\Delta| > \Lambda_\sigma^2$}}}: 
When $m_3 < m$, the parameter $\Delta$ is negative and the sigma model
potential has two discrete minima at $n_3=\pm 1$ corresponding to the
north and south poles respectively. These two degenerate vacua, are
clearly the ground states as long as $|\Delta|\gg \Lambda^2_\sigma$
and the sigma model coupling is weak. A choice of vacuum spontaneously
breaks the ${\mathbb Z}_2$ symmetry under $n_3\rightarrow -
n_3$. In this semiclassical regime the spectrum consists of massive
perturbative excitations and kinks that interpolate between the two
vacua. Since the model still has a $U(1)$ symmetry generated by
rotations in the 1-2 plane, the kink (and anti-kink) solutions have a
one-parameter degeneracy corresponding to this internal rotation
angle. One can then have solutions where this internal collective
coordinate is time-dependent and kinks rotate around the
$z$-axis. Bohr-Sommerfeld quantization of the semiclassical solution
implies that the associated conserved charge is quantized and the
sigma model kinks are ``dyonic'' \cite{qkinks}. The semiclassical
kinks (anti-kinks) can be labelled by the topological
kink number $T=+1$ (-1), and  the global $U(1)$ charge $S$. The mass
of the $(S,T)$ kink is \cite{qkinks}
\be
M_{S,T}^2=A_{k,N}\,\Delta\left[\tfrac{T^2}{4
    g_\sigma^4}+\left(S +\tfrac{\theta}{2\pi}T\right)^2\right].
\ee
The formula incoporates a 2D version of the Witten effect,
whereby a
non-zero vacuum angle induces a $U(1)$ charge,
$T
\tfrac
{\theta}{2\pi}$, for the kink.
As $\theta$ is smoothly varied from zero to $2\pi$, 
the semiclassical kink spectrum
undergoes a rearrangement. At $\theta=\pi$ there is a
level crossing, and the $(S,+1)$ and $(-S-1,+1)$ states
(and their charge conjugates) become degenerate.

The north and south pole vacua with $n_3=\pm 1$ correspond to two
different orientations of the non-Abelian magnetic flux. Taking the
flux in the $n_3=+1$ vacuum to be proportional to the matrix $Y_k$, we can
obtain the the $k$-string flux in the second vacuum by performing a
colour-flavour rotation in the 1-3 plane,
\be
e^{i\pi J_2} \,Y_k\,e^{-i\pi J_2} = - Y_{N-k}.\label{kinkcf}
\ee
Thus, in going from the north to the south pole, we do not change the
$N$-ality of the string, but we interpolate between a $k$-string and 
an anti-$(N-k)$-string. The interpolating kink carries zero $N$-ality
and is akin to a baryon vertex or a ``gluelump'' (Fig. 2).
\begin{figure}[h]
\begin{center}
\epsfig{file=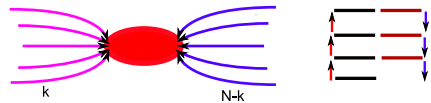, width =2.2in}
\end{center}
\caption{ The kink interpolating between the two sigma model vacua
  pictured as a ``gluelump'', and the level crossing as a function of $\theta$.}
\end{figure}\\
As $|\Delta|$ is decreased and approaches the dynamical scale
$\Lambda^2_\sigma$, the two classical vacua above mix quantum
mechanically, resulting in a
single global ground state, and a second local
minimum. The kinks and anti-kinks interpolate betwen these local minima, and 
are actually doublets of $SO(3)\simeq SU(2)$ \cite{witten}, and form
a stable bound state transforming as a massive triplet. Thus we expect
that the sigma model must undergo a phase transition in between the
two massive regimes dicussed above for $\Delta <0$. 
One can verify the presence of
such a phase transition easily, for the
anisotropic $O(n)$ model in the large-$n$ limit. In addition, in the
$O(3)$ model, at the special value
of $\theta=\pi$, for small enough $|\Delta|$, we expect two degenerate
vacua, with deconfined kinks and anti-kinks. At this point the model
is massless and 
the exact S-matrix between the $SU(2)$ doublets is also known
\cite{haldane}. 
\begin{figure}[h]
\begin{center}
\epsfig{file=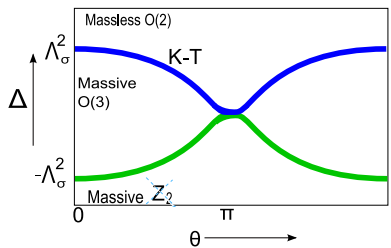, width =2.0in}
\end{center}
\caption{The phase plot of the $k$-string sigma model. 
The blue curve
  represents a K-T phase transition.}
\end{figure}

Based on the arguments above, we arrive at the phase diagram for
the sigma model on the $k$-string in Fig. 3.
It should be emphasized  that above 
the K-T phase transition 
the L\"uscher term \cite{luscher}
associated to the (magnetic) confining string in $D=4$
will jump from the value $(D-2)/24$ to $(D-1)/24$,
due to the contribution of the extra $O(2)$
massless degree of freedom.\\
{\it Kinks and Confined monopoles.--}
We will now attempt to establish a connection between the sigma model
dynamics above and 4D gauge theory physics, by looking at the regime
$\Delta<0$ but with $m_3\ll m$. We have seen that the
physics of the sigma model can change dramatically at $\theta=\pi$. From
\eqref{sigmamodel}, this
corresponds to the gauge theory vacuum angle taking the values 
$\theta_{YM}= \pi/k(N-k)$. It is {\it a priori} not clear what is the
significance of these values of $\theta_{YM}$. The answer to this
question will also reveal the connection between the kinks above and
confined monopoles, similar to the phenomena discovered in \cite{tong}.

For $m_3\ll m$, the ${\cal N}=1^*$ theory can be viewed as softly
broken ${\cal N}=2^*$ theory which is the theory of the ${\cal N}=2$
vector multiplet coupled to an adjoint hypermultiplet of mass $m$ \cite{dw}. 
The latter theory, with $m_3=0$, has a Coulomb branch
moduli space of vacua where the $SU(N)$ gauge group is broken to
$U(1)^{N-1}$. Singularities on the moduli space, 
where new light degrees of freedom appear
and in particular, where the Seiberg-Witten curve undergoes maximal
degeneration, descend to massive vacua  
of ${\cal N}=1^*$ theory upon soft-breaking with $m_3\neq0$. (It should
be emphasized that in this regime,  
the magnetic flux tubes are of the Abelian type and the sigma model
description will be inapplicable). The Higgs vacuum singularity 
is located at $\Phi_3=i m J_3$, $\Phi_{1,2}=0$. 
Denoting the diagonal
elements of the $\Phi_3$ VEV at this point as ${\bf a}= \{a_i\}$,
$i=1,\ldots, N$:
\bea
{\bf a} = m \,\left[\tfrac{(N-1)}{2}, \tfrac{(N-3)}{2},\ldots,
-\tfrac{(N-3)}{2}, -\tfrac{(N-1)}{2}\right]\,.
\eea
At weak coupling $g_{YM}\ll 1$, in this vacuum, the ${\cal N}=2^*$ theory
has massless electric hypermultiplets, 
and a semiclassical spectrum consisting of a
tower of BPS monopoles and dyons charged under the low-energy Abelian
groups.  The mass of a BPS state with magnetic and electric charges
$(n_m^i,n_e^i)$ is given by
\be
M=\sqrt{2}\left|{\bf a}\cdot({\bf n}_m\tau + {\bf n}_e)
+ m S \right|,\,\,\,\,
\tau \equiv \tfrac{4\pi i}{g^2_{YM}}+\tfrac{\theta_{YM}}{2\pi}.
\label{mass}
\ee
Crucially, the formula involves the charge $S$ for 
the global $U(1)$ symmetry of ${\cal N}=2^*$ theory which rotates
$\Phi_1$ and $\Phi_2$ into each other. This 
descends to the global $U(1)$ symmetry visible in the vortex sigma
model theory for $m_3 \lesssim m$. The BPS monopole
in the ${\cal N}=2^*$ theory, which carries the same magnetic charge
as the sigma model kink, has (from \eqref{kinkcf}) 
\be
{\bf n}_m = Y_{k}-(-Y_{N-k}) 
= (\underbrace{1,\ldots,1}_{k\,{\rm times}},0,0,\ldots,
\underbrace{-1,\ldots,-1}_{k\,{\rm times}}),\nonumber
\ee
and ${\bf n}_e=0$.
The fermion zero modes of the monopole due to the matter multiplet
$\Phi_{1,2}$ can be used to construct a multiplet with this magnetic
charge and non-zero $S$ \cite{ferrari}. 
The mass of such a state, using \eqref{mass}, is
$ M_S = {\sqrt 2} m\left|k(N-k)\tau + S\right|.$

It is clear from this formula that two such states with global
charge $S$ and $-S-1$, will become degenerate precisely when
$k(N-k)\theta_{YM}=\pi$. But we also know 
that for $m_3\neq 0$, in the vortex sigma model
$\theta=k(N-k)\theta_{YM}$ and that 
the (dyonic) kinks with $U(1)$ charge $S$ and $-S-1$ in the sigma
model undergo a level crossing at $\theta=\pi$.
Thus we are led to the conclusion that the
BPS monopoles above are confined on the $k$-strings when a non-zero
$m_3$ is introduced and then become identified with the kinks and
their ``dyonic'' excitations.

{\it Conclusions.--} We have shown that the quantum dynamics of certain
kinds of confining strings can exhibit novel phase transitions as parameters
of the underlying gauge theory are varied. The Kosterlitz-Thouless 
transition from a massive to massless phase manifests itself as a jump
in a physical observable, namely the L\"uscher term. 
It would be
interesting to find if such behaviour is generic to a wider class of
field theories, and what it says about the underlying gauge theory.
\\
{\bf Acknowledgements:} We would like to thank Nick Dorey for
discussions. 

\end{document}